\documentclass[12pt]{iopart}
\usepackage{iopams}
\begin{document}
\noindent \\
{\textbf{\Large{Tensor polarization of $\omega$ produced
at threshold in $p-p$ collisions}}
\author{G Ramachandran$^1$, J Balasubramanyam$^{2,3}$, S P Shilpashree$^1$
and G Padmanabha$^4$}
\address{$^1$ Indian Institute of Astrophysics, Koramangala, Bangalore-560 034,
India}
\address{$^2$ K. S. Institute of Technology, Bangalore-560 062, India}
\address{$^3$ Department of Physics, Bangalore University, Bangalore-560 056, India}
\address{$^4$ Sri Bhagawan Mahaveer Jain College, Bangalore-560 004, India}
\ead{gr@iiap.res.in}

\begin{abstract}
It is shown that the dominant decay mode of $\vec\omega \to \pi^+ \pi^- \pi^0$
can be employed to determine the Fano statistical tensor $t^2_0$ of
$\vec \omega$ with respect
to the quantization axis normal to the decay plane. In $pp \to pp \vec \omega$
one can choose decay planes with different orientations for a given
$\omega$ direction, $\theta$. By choosing three different experimentally convenient orientations of the decay planes for the same $\theta$, one may determine empirically the $t^2_0$ and $t^2_{\pm 2}$ characterising the tensor polarization of $\omega$ in the Transverse Frame for $pp \to pp \vec \omega$. 
\end{abstract}

\noindent \\
Meson production in $N-N$ collisions has attracted considerable attention
during the last decade of the past century as total cross section measurements
for pion production in the early years of the decade were found to be more
than a factor of 5 than the then available theoretical predictions. This 
catalyzed a variety of theoretical approaches and Hanhart \etal~\cite{1}
in 2000 have observed: `` As far as microscopic model calculations of the
reaction on $NN \to NN\pi$ are concerned one has to concede that theory is
definitely lagging behind the development of the experimental sector". With
measurements of spin observables in charged~\cite{2} as well as neutral~\cite{3}
pion production employing a polarized beam on a polarized target, experimental
studies have indeed reached a high degree of sophistication. Reference may be
made to several excellent reviews on these developments~\cite{4}. The 
J$\ddot{\rm u}$lich meson exchange model~\cite{1}, which yielded theoretical
predictions closer to data than most other models, has been more successful in
the case of charged pion production~\cite{2} than with neutral pions~\cite{3}.
A recent analysis~\cite{5} of $\vec p \vec p \to pp\pi^0$ measurements~\cite{3},
following a model independent irreducible tensor 
approach~\cite{6}, showed that the J$\ddot{\rm u}$lich model deviates from 
the empirically extracted estimates quite significantly for the
$^3P_1 \to\; ^3P_0 p$ and to a lesser extent for the $^3F_3 \to\; ^3P_2 p$ 
transitions; this
analysis has also been carried out with and without taking into consideration
the $\Delta$ contribution to emphasize its importance in the model calculation.
As the c.m. energy is increased further, thresholds for heavier meson
production are reached and one has to consider also contributions from excited
nucleon states. Since theoretical models for the nucleon
predict resonance states not seen
in $\pi-N$ scatterng, experimental studies on
meson production in $N-N$ collisions and on photo production 
of mesons~\cite{7} are useful to look for the
`missing resonances'~\cite{8} as well. 
It is also interesting to note that
production of isoscalar mesons like $\omega$ and $\phi$ involve only excited
nucleon states, in contrast to the case of isovector pion production.
Moreover, since meson production in
$N-N$ collisions probes short range hadron dynamics, it is particularly of
interest to study heavy meson production. While the
distance probed is $\approx $0.53 fm in the case of pion production, it goes down to
0.21 fm for $\omega$ and 0.18 fm for $\phi$ production~\cite{9}. In the 
absence of strange quarks in the initial state, the Okubo-Zweig-Iizuka (OZI)
rule~\cite{10} supresses $\phi$ production relative to $\omega$  production.
In view of the dramatic violation~\cite{11} of this rule observed in $\bar p p$
collisions, the ratio $R_{\phi/\omega}$ was measured~\cite{12} and was
found to be an order of magnitude larger, after correction for the available
phase space, than the theoretical estimate $R_{OZI} = 4.2 \times 10^{-3}$
~\cite{13}. The latest experimental estimate~\cite{14} is $R_{\phi/\omega} 
\approx 8 \times R_{OZI} $. 
Heavy meson production has also attracted attention in the
context of di-lepton spectra and medium modifications~\cite{15}. There is
a proposal~\cite{16} for the experimental study of heavy meson production 
in $\vec N \vec N$ collisions. Moreover the
state of polarization of vector meson
provides an interesting observable in addition to the differential
cross section and analyzing powers. Concerning ourselves here with $\omega$ 
production, the total cross section has been measured at five different
c.m. energies in the range 3.8 MeV to 30 MeV above threshold by Hibou \etal
~\cite{17} for $pp \to pp\omega$ and by Barsov \etal~\cite{18} for
$pn \to d\omega$. Measurements of the total cross section and angular
distribution have also been reported on $pp \to pp\omega$ at excess energies
of 92 MeV and 173 MeV~\cite{19} and at 60 MeV and 92 MeV~\cite{20} apart from
data~\cite{21} yet to be published. The
model independent irreducible tensor approach~\cite{6} has been extended to
$\omega$ production~\cite{22} where it was pointed out that the state of
polarization of $\omega$ can be studied experimentally using the decay mode
$\omega \to \pi^0 \gamma$. The importance of measuring the polarization of
$\omega$ has also been highlighted to determine empirically the 
threshold partial wave amplitudes~\cite{23} for $pp \to pp\omega$. It is learnt~\cite{24}
that it has been possible to fully reconstruct the decay of the
$\omega$ into three pions and identify the orientation of the decay plane with respect to 
the $\omega$ direction as well as the beam direction. Therefore, the purpose of
the present paper is to examine if the dominant decay mode of 
$\omega \to \pi^+ \pi^- \pi^0$
could be utilized to study the state of polarization of $\omega$.
\\
\\
Employing the irreducible tensor operators ${\bf S}^\lambda_\mu(s_f,s_i)$ 
of rank $\lambda$ connecting initial and final channel spin states $s_i$
and $s_f$ of hadrons as defined in~\cite{25} and following the 
approach~\cite{26} of Dalitz, we may
write~\cite{27}, the decay matrix for $\omega \to \pi^+ \pi^- \pi^0$ as
\begin{equation}
{\bf M}= f {\bf (S}^1(0,1). Q^1),
\end{equation}
where $f$ is a symmetric function of the energies of the 3 pions and 
$Q^1_\mu$ denote the spherical components of
\begin{equation}
{\bf Q  = [q_1 \times q_2 + q_2 \times q_3 + q_3
\times q_1]},
\end{equation}
in terms of the momenta $\bf q_1, q_2, q_3$ of the three pions which add up to
zero in the $\omega$ rest frame. We may define a right handed coordinate system
with z-axis along ${\bf Q}$ and x-axis along say ${\bf q_1}$ in the decay plane.
This may be referred to as the Decay Frame ($DF$).
Following the Madison convention~\cite{28}, 
the spin density matrix ${\brho}$ of a polarized $\omega$ may be written in the form
\begin{equation}
{\brho} = {\Tr {\brho} \over 3} [ 1+ \sum_{k=1}^2 ({\btau^k }\cdot t^k)],
\end{equation}
in terms of the standard $3\times 3$ matrices ${\btau^k_\mu}, \mu = -k,..,k$ and the
Fano statistical tensors $t^k_\mu$ of rank $k$, characterising the polarized
$\omega$. Noting that ${\btau}^k_\mu = {\bf S}^k_\mu(1,1)$ and 
making use of the known~\cite{25}
properties of the irreducible tensor operators along with Racah techniques, the 
angular distribution of the three pions in the decay plane is given by
\begin{equation}
{\bf M} {\brho} {\bf M^\dagger} = |f|^2{\Tr {\brho} \over 3} [{(\bf Q \cdot Q)}-
\sqrt 3 \sum_{k=1}^2
(t^k \cdot (Q^1 \otimes Q^1)^k)],
\end{equation}
where ${\bf M^\dagger}$ denotes the hermitian conjugate of ${\bf M}$. The first term ${(\bf Q \cdot Q)}$ leads to the well-known Dalitz plot for the $3\pi$ decay
of unpolarized $\omega$. 
Since $(Q^1 \otimes Q^1)^1_\mu = 0$, the $k=1$ term drops out, so that (4) 
contains only the tensor analyzing power
\begin{equation}
A^2_0(DF)= -\sqrt 3 (Q^1 \otimes Q^1)^2_0,
\end{equation}
since $(Q^1 \otimes Q^1)^2_\mu= \delta_{\mu 0} {\sqrt{2 \over 3}}
({\bf Q \cdot Q}) $ in $DF$.
We have
\begin{equation}
{\bf M} {\brho} {\bf M^\dagger} = {|f|^2 \Tr {\brho} \over 3} ({\bf Q \cdot Q})[1- \sqrt 2 t^2_0(DF)],
\end{equation}
where $t^2_0(DF)$ denotes the $\mu = 0$ component of the tensor polarization
$t^2_\mu$ of the $\vec \omega$ in its decay frame, $DF$. Thus the distribution of 
points in the Dalitz plot retains the same shape, even when $t^2_0(DF)\neq 0$.
However, since  $-\sqrt 2 \leq t^2_0 \leq {1 \over \sqrt 2}$, the total number of points N$(DF)$ in the associated decay  plane is sensitive to the
tensor polarization $t^2_0(DF)$ of $\vec \omega$ as the multiplicative factor $(1- \sqrt 2
t^2_0(DF))$ may vary between 0 and 3. 
\\
In the case of $\vec \omega$
in $pp \to pp \vec \omega$, we may identify $ \Tr {\brho}$ with the
unpolarized differential cross section ${ d\sigma_0 \over d\Omega}$, which is
given as a function of the angle $\theta$ between the momentum ${\bf q}$ of the
$\omega$ and the momentum ${\bf p_i}$ of the proton beam in the c.m. frame for
the reaction. A right handed Cartesian coordinate system with
z-axis along ${\bf p_i \times q}$ and x-axis along ${\bf p_i }$, was
referred to~\cite{23} as the Transverse Frame($TF$), where in the Fano statistical tensors
$t^2_\mu$ characterising the tensor polarization of the $\vec \omega$ 
are such that $t^2_{\pm 1}=0$ and
\begin{eqnarray}
\eqalign{
\Tr {\brho} \hspace{1mm} t^2_0 &= {1\over 384 \pi^2} \sqrt{1\over 2} \int dW\, \big(
[0.6|f'|^2-|f_1|^2] \cr
& \qquad
+1.8 \cos^2\theta[10|f_2|^2+2|f_3|^2-|f'|^2]\big),}
\end{eqnarray}
\begin{eqnarray}
\eqalign{
\Tr {\brho} \hspace{1mm} t^2_{\pm 2} &= {1\over 256 \pi^2} \sqrt{1\over 3} \int dW\, \big(
[|f_1|^2+0.6|f'|^2]-1.2\cos^2\theta[15|f_2|^2  \cr & \qquad 
+3|f_3|^2   -0.5|f'|^2]
\mp 0.6i \sin 2\theta[15|f_2|^2-1.5|f_3|^2-0.5|f'|^2]\big),}
\end{eqnarray}
while
\begin{eqnarray}
\Tr {\brho}= {d \sigma_0 \over d\Omega} &= {1\over 192 \pi^2} \int dW\, \big(
[|f_1|^2+0.3|f'|^2] \cr & \qquad +0.9 \cos^2\theta[10|f_2|^2+2|f_3|^2-|f'|^2]\big)
\end{eqnarray}
in terms of the lowest three partial wave amplitudes $f_1, f_2, f_3$ for 
$pp \to pp \omega$ given in Table 1 of ~\cite{23} and $f' = \sqrt 10 f_2 +
f_3$. The integration with respect to the invariant mass $W$ of the two 
protons in the final state may also be replaced by integration with respect to the c.m.
energy $E_\omega$ of $\omega$ using eqn(1) of~\cite{23}. If $(\alpha, \beta,
\gamma)$ denote the Euler angles of rotation for going from $TF$ to $DF$, we have
\begin{equation}
t^2_0(DF) = \sum_{\mu=-2}^2 D^2_{\mu o} (\alpha, \beta, \gamma) t^2_\mu.
\end{equation}
Thus $t^2_0(DF)$ depends not only on the
angle $\theta$ at which the $\vec \omega$ is produced, 
but also on the Euler
angles $(\alpha, \beta, \gamma)$ which characterise the orientation of its
decay plane with respect to $TF$. Counting N$(DF)$ in a given decay 
plane provides, by virtue of (6), an estimate 
of the associated $t^2_0(DF)$. Noting that $t^2_{-2}$ is the complex conjugate of $t^2_{2}$, it is clear that eqn.(10) is a linear equation in three real unknowns viz., $t^2_0, Re(t^2_2)$
and $Im(t^2_2)$ in the $TF$. One may conveniently choose three different sets of $(\alpha,
\beta,\gamma)$ to measure experimentally the corresponding real numbers, $t^2_0(DF)$ on the
left hand side of eqn.(10) for $\vec \omega$ produced in the same direction $\theta$. Having set up an appropriate set of three linear equations, one may then solve for $t^2_0, Re(t^2_2)$ and $Im(t^2_2)$ in the $TF$ which  characterise completely the tensor polarization of $\omega$ produced in any chosen direction.

\ack{
We thank Professor K. T. Brinkmann for encouraging us to look into this problem.
GR and SPS thank Professors B. V. Sreekantan , R. Cowsik, J. H. Sastry,
R. Srinivasan and S. S. Hasan
for the facilities provided for research at the Indian Institute
of Astrophysics. JB and GP thank Principal and Management of their respective
organizations for encouragement.
}


\section*{ References}
\begin{thebibliography}{}
\bibitem{1} Hanhart C, Haidenbauer J,  Krehl O and Speth J 2000
{\it Phys. Rev.} C {\bf 61}, 064008 
\bibitem{2} Von Prezewoski B \etal 2000 {\it Phys. Rev.} C {\bf 61} 064604
\\Daehnick W W \etal 2002 {\it Phys. Rev.} C {\bf 65} 024003
\bibitem{3} Meyer H O \etal 2001 {\it Phys. Rev.} C {\bf 63} 064002
\bibitem{4} Machner H and Haidenbauer J 1999 {\it J. Phys. G: Nucl. Part. Phys.}
{\bf 25} R231\\
Moskal P, Wolke M, Khoukaz A and Oelert W 2002 {\it Prog. Part. Nucl. Phys.}
{\bf 49} 1\\
F\"aldt G, Johnsson  T and Wilkin C 2002 {\it Physica Scripta T} {\bf 99}
146
\\Hanhart C 2004 {\it Phys. Rep.} {\bf 397} 155 
\bibitem{5} Deepak P N, Haidenbauer J and Hanhart C 2005 {\it Phys. Rev.} C
{\bf 72} 024004
\bibitem{6} Ramachandran G, Deepak P N and Vidya M S 2000 {\it Phys. Rev.} C
{\bf 62} 011001(R) 
\\Ramachandran G and Deepak P N 2000 {\it J. Phys. G: Nucl. Part. Phys.}
{\bf 26} 1809
\\Ramachandran G and Deepak P N 2001 {\it Phys. Rev.} C {\bf 63} 051001(R)
\\Deepak P N and  Ramachandran G 2002 {\it Phys. Rev.} C {\bf 65} 027601
\\Deepak P N, Hanhart C, Ramachandran G and Vidya M S 2005
{\it Int. J. Mod. Phys.} A {\bf 20} 599
\bibitem{7} Krusche V and Schadmand S
2003 {\it Prog. Part. Nucl. Phys.} {\bf 51} 399
\\Burkert V D and Lee T S H
2004 {\it Int. J. Mod. Phys.} E {\bf 13} 1035 
\bibitem{8}Barns T and Morsch H P (Eds) 2000 Baryon
Excitations,  Lectures of COSY Workshop held at Forschungszentrum
J\"ulich, 2-3 May 2000, ISBN 3-89336-273-8
\\
Carlson C and Mecking B 2003 International Conference on
the structure of Baryons,  BARYONS 2003, New Port News,
Virginia 3-8 March 2002 (World Scientific, Singapore).
\\
Dytman S A and Swanson E S 2003
Proceedings of NSTAR 2002 Wokshop, Pittsburg 9-12 Oct
2003 (World Scientific, Singapore).
\bibitem{9} K. Nakayama 2002 Proc. Symposium on ``Threshold Meson Production in
$pp$ and $pd$ Interactions", Schriften des Forschungszentrum J\"ulich,
{\it Matter.Mater}, {\bf{11}} 119 
\bibitem{10}  Okubo S 1963 {\it Phys. Lett.  } B{\bf 5} 165
\\Zweig G 1964 {\it CERN Report} 8419/TH412 
\\Iizuka I 1966 {\it Prog. Theor. Phys. Suppl.} {\bf{37-38}} 21
\bibitem{11} Amsler C 1998 {\it Rev. Mod. Phys}. {\bf 70} 1293
\bibitem{12} Balestra F \etal 1998 {\it Phys. Rev. Lett} {\bf 81} 4572
\\ Balestra F \etal 2001  {\it Phys. Rev.} C {\bf 63} 024004
\bibitem{13} Lipkin H J 1976 {\it Phys. Lett} B {\bf 60} 371
\bibitem{14} Hartmann M \etal 2006 {\it Phys. Rev. Lett} {\bf 96} 242301
\bibitem{15} Jain B K and Santra A B 1993 {\it Phys. Rep} {\bf 230} 1
\\Faessler A, Fuchs C and Krivoruchenko M I 2000 {\it Phys. Rev.}
C {\bf 61} 035206
\\Fuchs C, Faessler A, Cozma D, Martemyanov B V and  Krivoruchenko M I
 {\it Nucl. Phys.} A {\bf 755} 499
\bibitem{16} Rathmann F \etal 1999 Study of Heavy Meson production in $NN$
collisions with polarized Beam and Target, Letter of Intent to COSY-PAC No.
81 (J\"ulich)
\\ Rathmann F \etal 2002 {\it Czec. J. phys}, {\bf 52}, c319 
\bibitem{17} Hibou F \etal 1999 {\it Phys. Rev. Lett}. {\bf{83}} 492 
\bibitem{18} Barsov S \etal 2004
{\it Eur. Phys. J.} A {\bf 21} 521
\bibitem{19} Abd El-Samad  S \etal (COSY-TOF Coll) 2001 {\it Phys. Lett.} B
{\bf 522} 16
\bibitem{20} Barsov S \etal 2006
arXiv: nucl-ex/0609010 v1 8 Sep 2006
\bibitem{21} Martin Schulte-Wissermann 2004 Doctoral Dissertation(Unpublished),
Technschen Universitat, Dresden 
\bibitem{22} Ramachandran G, Vidya M S, Deepak P N, Balasubramanyam J
and Venkataraya 2005
{\it Phys. Rev.} C {\bf 72} 031001(R)
\bibitem{23} Ramachandran G, Balasubramanyam J, Vidya M S and Venkataraya 2006
{\it Mod. Phys. Lett.} A  {\bf 21} 2009
\bibitem{24} Brinkmann K T 2006 {\it private communication}
\bibitem{25} Ramachandran G and Vidya M S 1997 {\it Phys. Rev.} C {\bf 56} R12
\bibitem{26} Dalitz R H 1953 {\it Phil Mag.} {\bf 44} 1068
\\Dalitz R H 1954 {\it Phys. Rev.} {\bf 94} 1046
\\Fabri E 1954 {\it Nuovo cimento} {\bf 11} 479
\\Dalitz R H 1957 {\it Rep. Prog. Phys.} {\bf 20} 163
\\Stevenson M L \etal 1962 {Phys. Rev.} {\bf 125} 687
\\Charles Zemach 1964 {\it Phys. Rev.} {\bf 133} B1201
\bibitem{27} K$\ddot{\rm a}$ll\'en G 1964 {\it Elementary Particle Physics}
(Addison-Wesley Publishing Co) p 201
\bibitem{28} Satchler G R \etal 1970 {\it Proc 3rd Int. Symp. on
Polarization
Phenomena in Nuclear Reactions} Eds Barschall H H and Haeberli W(Eds) (Madison, WI:
University of Wisconsin Press) p XXV
\bibitem{29}  Rose M E 1957{\it Elementary Theory of Angular momentum}
(New York: John Wiley)

\end{thebibliography}
\end{document}